\documentclass{article}
\usepackage{amssymb}

\usepackage{latexsym}
\usepackage{amsmath}
\usepackage{graphicx}
\usepackage{float}

\newlength{\dinwidth}
\newlength{\dinmargin}
\begin{document}



\thispagestyle{empty} \vspace*{1cm} 
\vspace*{2cm}

\begin{center}
{\LARGE Twisted Conformal Field Theories and Morita equivalence }

{\LARGE \ }

{\large Vincenzo Marotta\footnote{{\large {\footnotesize Dipartimento di
Scienze Fisiche,}{\it \ {\footnotesize Universit\'{a} di Napoli ``Federico
II''\ \newline
and INFN, Sezione di Napoli}, }{\small Compl.\ universitario M. Sant'Angelo,
Via Cinthia, 80126 Napoli, Italy}}},} {\large Adele Naddeo\footnote{{\large
{\footnotesize CNISM, Unit\`{a} di Ricerca di Salerno and Dipartimento di
Fisica {\it ''}E. R. Caianiello'',}{\it \ {\footnotesize Universit\'{a}
degli Studi di Salerno, }}{\small Via Salvador Allende, 84081 Baronissi
(SA), Italy}}}\footnote{{\large {\footnotesize Dipartimento di Scienze
Fisiche,}{\it \ {\footnotesize Universit\'{a} di Napoli ``Federico II''}, }%
{\small Compl.\ universitario M. Sant'Angelo, Via Cinthia, 80126 Napoli,
Italy}}}}

{\small \ }

{\bf Abstract\\[0pt]
}
\end{center}

\begin{quotation}
The Morita equivalence for field theories on noncommutative two-tori is
analysed in detail for rational values of the noncommutativity parameter $%
\theta $ (in appropriate units): an isomorphism is established
between an abelian noncommutative field theory (NCFT) and a
non-abelian theory of twisted fields on ordinary space. We focus
on a particular conformal field theory (CFT), the one obtained by
means of the $m$-reduction procedure \cite {VM}, and show that it
is the Morita equivalent of a NCFT. Finally, the whole
$m$-reduction procedure is shown to be the image in the ordinary
space of the Morita duality. An application to the physics of a
quantum Hall fluid at Jain fillings $\nu =\frac{m}{2pm+1}$ is
explicitly discussed in order to further elucidate such a
correspondence and to clarify its role in the physics of strongly
correlated systems. A new picture emerges, which is very different
from the existing relationships between noncommutativity and many
body systems \cite{ncmanybody}.

\vspace*{0.5cm}


{\footnotesize PACS: 11.25.Hf, 11.10.Nx, 03.65.Fd\newpage }\baselineskip%
=18pt \setcounter{page}{2}
\end{quotation}

\section{Introduction}

Noncommutative field theories (NCFT) have attracted much attention in the
last years because they provide a non trivial generalization of local
quantum field theories, allowing for some degree of non locality while
retaining an interesting mathematical structure \cite{ncft1}. Really they
should be viewed as living somewhere between ordinary field theory and
string theory. Indeed they naturally arise as some low energy limit of open
string theory and as the compactification of $M$-theory on the torus \cite
{ncft2}. Space-time noncommutativity also arises naturally when the dynamics
of open strings attached to a $D2$-brane in a $B$ field background is
considered \cite{ncft3}: in such a case the open strings act as dipoles of $%
U\left( 1\right) $ gauge field of the brane and their scattering amplitudes
in the low energy limit are properly described by a Super Yang-Mills gauge
theory defined on a noncommutative two-torus with deformation parameter $%
\theta $ identified with the\ $B$ field \cite{ncft4}. More generally, gauge
theories on tori with magnetic flux and twisted models can be reformulated
in terms of noncommutative gauge theories. One motivation for the relevance
of such theories is that the notion of space-time presumably has to be
modified at very short distances so that the effect of the granularity of
the space can be taken into account.

A significant feature of NCFT is the celebrated Morita duality
between noncommutative tori \cite{morita}. This duality is a
powerful mathematical result that establishes a relation, via an
isomorphism, between two noncommutative algebras. In principle it
was introduced as a resolution to certain paradoxes which arise in
the context of the reconstruction of topological spaces from
$C^{\ast }$-algebras \cite{morita}. Of particular importance are
the algebras defined on the noncommutative torus, where it can be
shown that Morita equivalence holds if the corresponding sizes of
the tori and the noncommutative parameters are related in a
specific way \cite {torus1}. Several results have been established
in the literature about the Morita equivalence of NCFT but
principally focused on noncommutative gauge theories and
describing mostly classical or semiclassical aspects of them
\cite{morita1}. Indeed Morita duality of gauge theories on
noncommutative tori is a low energy analogue of $T$-duality of the
underlying string model \cite{string1}\cite{string2}; when
combined with the hypothesis of analiticity as a function of the
noncommutativity parameter, it gives information about singular
large-$N$ limits of ordinary $U(N)$ gauge theories
\cite{alvarez1}. Nevertheless another point of view can be
usefully developed in order to establish a correspondence between
NCFT and well known standard field theories. Indeed, for special
values of the noncommutativity parameter, one of the isomorphic
theories obtained by using the Morita equivalence is a commutative
field theory on an ordinary space
\cite{moreno1}\cite{moreno2}\cite{schiappa}.

Here we follow this line and concentrate on a particular conformal
field theory (CFT), the one obtained via $m$-reduction technique
\cite{VM}, which has been recently applied to the description of a
quantum Hall fluid (QHF) at Jain \cite{cgm1}\cite{cgm3} as well as
paired states fillings \cite{cgm2}\cite{cgm4} and in the presence
of topological defects \cite{noi1}\cite{noi2}\cite{noi5}; by using
Morita duality we build up the essential ingredients of the
corresponding NCFT. In such a context the granularity of the space
is due to the existence of a minimum area which is the result of a
fractionalized magnetic flux. The $m$-reduction technique is based
on the simple observation that, for any CFT (mother), a class of
sub-theories exists, which is parameterized by an integer $m$ with
the same symmetry but different representations. The resulting
theory (daughter), called Twisted
Model (TM), has the same algebraic structure but a different central charge $%
c_{m}=mc$. Its application to the physics of the QHF arises by the
incompressibility of the Hall fluid droplet at the plateaux, which
implies its invariance under the $W_{1+\infty }$ algebra at
different fillings \cite {ctz5}, and by the peculiarity of the
$m$-reduction procedure to provide a daughter CFT with the same
$W_{1+\infty }$ invariance property of the mother theory
\cite{cgm1}\cite{cgm3}. Thus the $m$-reduction furnishes
automatically a mapping between different incompressible plateaux
of the QHF. The characteristics of the daughter theory is the
presence of twisted boundary conditions on the fundamental fields,
which coincide with the conditions required by the Morita
equivalence for a class of NCFT. Here the noncommutativity of the
spatial coordinates appears as a consequence of the twisting. As a
result, the $m$-reduction technique becomes the image in the
ordinary space of the Morita duality. Furthermore the Moyal
algebra, which characterizes the NCFT, has a natural realization
in terms of Generalized Magnetic Translations (GMT) within the
$m$-reduced theory when we refer to the description of a QHF at
Jain fillings $\nu =\frac{m}{2pm+1}$ \cite{cgm1}\cite{cgm3}. In
this paper we show how the $m$-reduction procedure induces a CFT
which is the Morita equivalent of a NCFT by making explicit
reference to the physics of a QHF at Jain fillings. We point out
that in the last years there have been many investigations on the
relationship between noncommutative spaces and QHF: in all such
studies noncommutativity is related to the finite number $N_{e}$
of electrons in a
realistic sample via the rational parameter $\theta \propto \frac{1}{N_{e}}$%
, which sets the elementary area of nonlocality \cite{qhf1}\cite{qhf2}\cite
{qhf3}\cite{qhf4}. In this context the $\theta $-dependence of physical
quantities is expected to be analytic near $\theta =0$ \cite{alvarez1}
because of the very smooth ultraviolet behaviour of noncommutative
Chern-Simons theories \cite{uv1}; as a consequence the effects of the
electron's granularity embodied in $\theta $ can be expanded in powers of $%
\theta $ via the Seiberg-Witten map \cite{ncft2} and then appear as
corrections to the large-$N$ results of field theory. Our approach is quite
different: the noncommutativity is present at field theory level and
constrains the structure of the theory already in the large-$N$ limit. It
can be ascribed to a residual noncommutativity for large numbers of
electrons which group in clusters of finite $m$. That introduces a new
relationship between noncommutative spaces and QHF and paves the way for
further investigations on the role of noncommutativity in the physics of
general strongly correlated many body systems \cite{ncmanybody}.

The paper is organized as follows.

In Section 2, we review the main steps to be performed in order to get a $m$%
-reduced CFT on the plane \cite{VM} and then we briefly recall the
description of a QHF at Jain fillings $\nu =\frac{m}{2pm+1}$ as a result of
the $m$-reduction procedure \cite{cgm1}\cite{cgm3}.

In Section 3, we explicitly build up the Morita equivalence between CFTs in
correspondence of rational values of the noncommutativity parameter $\theta $
with an explicit reference to the $m$-reduced theory describing a QHF at
Jain fillings. Indeed we clearly show that there is a well defined
isomorphism between the fields on a noncommutative torus and those of a
non-abelian field theory on an ordinary space.

In Section 4, we show in whole generality how the possible states of a QHF
on a discrete two-torus can be classified in terms of the coset $SL(2,Z_{m})$%
; in this way we underline the relevance of the hidden $su\left( m\right) $
algebra in classifying the topological excitations of the QHF.

In Section 5, we further clarify the deep relationship between $m$-reduction
procedure and Morita equivalence by showing how the NCFT properties can be
obtained from a generalization of the ordinary magnetic translations in a
QHF context \cite{cristofano1}.

In Section 6, some comments and outlooks of this work are given.

\section{The $m$-reduction procedure}

In this Section we review the basics of the $m$-reduction procedure on the
plane (genus $g=0$) \cite{VM} and then we show briefly how it works,
referring to the description of a QHF at Jain fillings $\nu =\frac{m}{2pm+1}$
\cite{cgm1}\cite{cgm3}.

In general, the $m$-reduction technique is based on the simple observation
that for any CFT (mother) exists a class of sub-theories parameterized by an
integer $m$ with the same symmetry but different representations. The
resulting theory (daughter) has the same algebraic structure but a different
central charge $c_{m}=mc$. In order to obtain the generators of the algebra
in the new theory we need to extract the modes which are divided by the
integer $m$. These can be used to reconstruct the primary fields of the
daughter CFT. This technique can be generalized and applied to any extended
chiral algebra which includes the Virasoro one. Following this line one can
generate a large class of CFTs with the same extended symmetry but with
different central extensions. It can be applied in particular to describe
the full class of Wess-Zumino-Witten (WZW) models with symmetry $\widehat{%
su(2)}_{m}$, obtaining the associated parafermions in a natural way or the
incompressible $W_{1+\infty }$ minimal models \cite{ctz5} with central
charge $c=m$. Indeed the $m$-reduction preserves the commutation relations
between the algebra generators but modify the central extension (i.e. the
level for the WZW models). In particular this implies that the number of the
primary fields gets modified.

The general characteristics of the daughter theory is the presence of
twisted boundary conditions (TBC) which are induced on the component fields
and are the signature of an interaction with a localized topological defect.
It is illuminating to give a geometric interpretation of that in terms of
the covering on a $m$-sheeted surface or complex curve with branch-cuts, see
for instance Figs. 1, 2 for the particular case $m=2$.

\begin{figure}[h]
\centering\includegraphics*[width=0.3\linewidth]{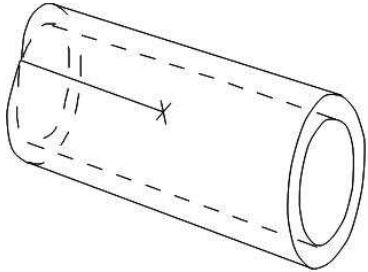}
\caption{The edge of the $2$-covered cylinder can be viewed as a separation
line of two different domains of the $2$-reduced CFT.}
\label{figura1}
\end{figure}

Indeed the fields which are defined on the left domain of the boundary have
TBC while the fields defined on the right one have periodic boundary
conditions (PBC). When we generalize the construction to a Riemann surface
of genus $g=1$, i. e. a torus, we find different sectors corresponding to
different boundary conditions on the cylinder, as shown in detail in Refs.
\cite{cgm3}\cite{cgm4}. Finally we recognize the daughter theory as an
orbifold of the usual CFT describing the QHF at a given plateau.

\begin{figure}[h]
\centering\includegraphics*[width=0.3\linewidth]{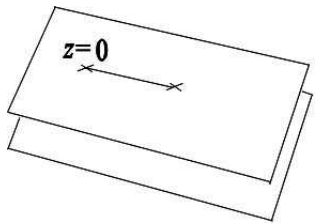}
\caption{The local branched plane.}
\end{figure}

The physical interpretation of such a construction within the context of a
QHF description is the following. The two sheets simulate a two-layer
quantum Hall system and the branch cut represents TBC which emerge from the
interaction with a localized topological defect on the edge \cite{noi1}\cite
{noi2}\cite{noi5}. Let us also notice that the $m$-reduction procedure was
first applied to 2D quantum gravity as a way to reinforce the so called
string constraints for the solutions \cite{FKN}.

Let us now briefly summarize the $m$-reduction procedure on the plane \cite
{VM}, which has been recently applied to the description of a quantum Hall
fluid (QHF) at Jain \cite{cgm1} as well as non standard fillings \cite{cgm2}%
. Its generalization to the torus topology has been given in Refs. \cite
{cgm3}\cite{cgm4}. The starting point is described by a CFT with $c=1$, in
terms of a scalar chiral field compactified on a circle with general radius $%
R^{2}$ ($R^{2}=1$ for the Jain series \cite{cgm1} while $R^{2}=2$ for the
non standard one \cite{cgm2}). Then the $u(1)$ current is given by $%
J(z)=i\partial _{z}Q(z)$, where $Q(z)$ is the compactified Fubini field with
the standard mode expansion:
\begin{equation}
Q(z)=q-i\,p\,lnz+\sum_{n\neq 0}\frac{a_{n}}{n}z^{-n},  \label{modes}
\end{equation}
where $a_{n}$, $q$ and $p$ satisfy the commutation relations $\left[
a_{n},a_{n^{\prime }}\right] =n\delta _{n,n^{\prime }}$ and $\left[ q,p%
\right] =i$. The primary fields are expressed in terms of the vertex
operators $U^{\alpha _{s}}(z)=:e^{i\alpha _{s}Q(z)}:$ with $\alpha _{s}=%
\frac{s}{R}$\ ($s=1,...,R^{2}$) and conformal dimension $h=\frac{s^{2}}{%
2R^{2}}$.

Starting with the set of fields in the above CFT and using the $m$-reduction
procedure, which consists in considering the subalgebra generated only by
the modes in Eq. (\ref{modes}), which are multiple of an integer $m$, we get
the image of the twisted sector of a $c=m$ orbifold CFT (i. e. the TM),
which describes the Lowest Landau Level (LLL) dynamics of the new filling in
the QHF context. In this way the fundamental fields are mapped into $m$
twisted fields which are related by a discrete abelian group. Indeed the
fields in the mother CFT can be factorized into irreducible orbits of the
discrete $Z_{m}$ group, which is a symmetry of the TM, and can be organized
into components, which have well defined transformation properties under
this group. To compare the orbifold so built with the $c=m$ CFT, we use the
mapping $z\rightarrow z^{1/m}$ and the isomorphism, defined in Ref. \cite{VM}%
, between fields on the $z$ plane and fields on the $z^{m}$ covering plane
given by the following identifications: $a_{nm+l}\longrightarrow \sqrt{m}%
a_{n+l/m}$, $q\longrightarrow \frac{1}{\sqrt{m}}q$.

We perform a \textquotedblleft double\textquotedblright\ $m$-reduction which
consists in applying this technique into two steps.

{\bf 1)} The $m$-reduction is applied to the Fubini field $Q(z).$ That
induces twisted boundary conditions on the currents. It is useful to define
the invariant scalar field:
\begin{equation}
X(z)=\frac{1}{m}\sum_{j=1}^{m}Q(\varepsilon ^{j}z),  \label{X}
\end{equation}
where $\varepsilon ^{j}=e^{i\frac{2\pi j}{m}}$, corresponding to a
compactified boson on a circle with radius now equal to $R_{X}^{2}=R^{2}/m$.
This field describes the $U(1)$ electrically charged component of the new
filling in a QHF description.

On the other hand the non-invariant fields defined by
\begin{equation}
\phi ^{j}(z)=Q(\varepsilon ^{j}z)-X(z),~~~~~~~~~~~~~~~~\sum_{j=1}^{m}\phi
^{j}(z)=0  \label{phi}
\end{equation}
naturally satisfy twisted boundary conditions, so that the $J(z)$ current of
the mother theory decomposes into a charged current given by $J(z)=i\partial
_{z}X(z)$ and $m-1$ neutral ones $\partial _{z}\phi ^{j}(z)$ \cite{cgm1}.

{\bf 2)} The $m$-reduction applied to the vertex operators $U^{\alpha
_{s}}(z)$ of the mother theory also induces twisted boundary conditions on
the vertex operators of the daughter CFT. The discrete group used in this
case is just the $m$-ality group which selects the neutral modes with a
complementary cut singularity, which is necessary to reinforce the locality
constraint.

The vertex operator in the mother theory can be factorized into a vertex
that depends only on the $X(z)$ field:
\begin{equation}
{\cal U}^{\alpha _{s}}(z)=z^{\frac{\alpha _{s}^{2}(m-1)}{m}}:e^{i\alpha _{s}{%
\ }X(z)}:
\end{equation}
and in vertex operators depending on the $\phi ^{j}(z)$ fields. It is useful
to introduce the neutral component:
\begin{equation}
\psi _{1}(z)=\frac{z^{\frac{1-m}{m}}}{m}\sum_{j=1}^{m}\varepsilon
^{j}:e^{i\phi ^{j}(z)}:
\end{equation}
which is invariant under the twist group given in {\bf 1)} and has $m$-ality
charge $l=1$. Then, the new primary fields are the composite vertex
operators $V^{\alpha _{s}}(z)={\cal U}^{\alpha _{s}}(z)\psi _{l}(z)$, where $%
\psi _{l}$ are the neutral operators with $m$-ality charge $l$.

\bigskip From these primary fields we can obtain the new Virasoro algebra
with central charge $c=m$ which is generated by the energy-momentum tensor $%
T(z)$. It is the sum of two independent operators, one depending on the
charged sector:
\begin{equation}
T_{X}(z)=-\frac{1}{2}:\left( \partial _{z}X\left( z\right) \right) ^{2}:
\label{VIR1}
\end{equation}
with $c=1$ and the other given in terms of the $Z_{m}$ twisted bosons $\phi
^{j}(z)$:
\begin{equation}
T_{\phi }(z)=-\frac{1}{2}\sum_{j,j^{\prime }=1}^{m}:\partial _{z}\phi
^{j}(z)\partial _{z}\phi ^{j^{\prime }}(z):+\ \frac{m^{2}-1}{24mz^{2}}
\label{VIR2}
\end{equation}
with $c=m-1$.

Let us notice here that, although the daughter CFT has the same central
charge value, it differs in the symmetry properties and in the spectrum,
depending on the mother theory we are considering (i.e. for Jain or non
standard series in the case of a QHF).

Let us now focus on the description of a QHF at Jain fillings in terms of
vertex operators and review the main results of $m$-reduction procedure in
order to classify its excitations. The starting point is a CFT with $c=1$,
in terms of a scalar chiral field compactified on a circle with radius $%
R^{2}=1$. Then the $U(1)$ current is given by $J(z)=i\partial _{z}Q(z)$,
where $Q(z)$ is the compactified Fubini field given in Eq. (\ref{modes}).
The primary fields are expressed in terms of the vertex operators $U^{\alpha
_{s}}(z)=:e^{i\alpha _{s}Q(z)}:$ with $s=1$ and conformal dimension $h=\frac{%
1}{2}$. The dynamical symmetry is given by the $W_{1+\infty }$ algebra \cite
{BS} with $c=1$, whose generators are simply given by a power of the current
$J(z)$. By using the $m$-reduction procedure, we get the image of the
twisted sector of a $c=m$ orbifold CFT which has $\widehat{U}(1){\times }%
\widehat{SU}(m)_{1}$ as extended symmetry and describes the QHF at the new
general filling $\nu =\frac{m}{2pm+1}$. In order to do so, we factorize the
fields into two parts, the first is the $c_{X}=1$ charged sector with radius
$R_{X}^{2}=\frac{2pm+1}{m}$, the second describes neutral excitations with
total conformal central charge $c_{\phi }=m-1$ for any $p\in N$ \cite{cgm1}.

In order to obtain a pure holomorphic wave function we have to consider the
correlator of the TM primary fields, which are the composite vertex
operators $V^{\alpha _{s}}(z)={\cal U}^{\alpha _{s}}(z)\psi _{l}(z)$
\footnote{$\psi _{l}$ are the neutral operators associated with
representations of $m$-ality $l$ of \ $\widehat{SU}(m)_{1}$\cite{cgm1}.}
with conformal dimension:
\begin{equation}
h_{l}=\frac{l^{2}}{2m\left( 2pm+1\right) }+\frac{a}{2}\left( \frac{m-a}{m}%
\right) ,\text{ \ \ \ \ \ \ \ \ }l=1,2,...,m\left( 2pm+1\right) ;
\label{cdim1}
\end{equation}
they describe excitations with electric charge $q_{e}=\frac{l}{2pm+1}$ and
magnetic charge $q_{m}=l$ in units of $\frac{hc}{e}$. There exist also
integer charge quasi-particles (termed $a$-electrons), with half integer (or
integer) conformal dimension given by:
\begin{equation}
h_{l}=a^{2}p+\frac{a}{2},\text{ \ \ \ \ \ \ \ \ }l=\left( 2pm+1\right) a;%
\text{ \ \ \ \ \ \ \ }a=1,2,...,m.  \label{cdim2}
\end{equation}
In particular the electrons are obtained in correspondence of $q_{e}=1$ and $%
q_{m}=2pm+1$, while the other $2pm$ primary fields correspond to anyons.

The spectrum just obtained follows from the construction of the Virasoro
algebra with central charge $c=m$ (see Eqs. (\ref{VIR1})-(\ref{VIR2})). We
should point out that $m$-ality in the neutral sector is coupled to the
charged one exactly, as it was derived in Refs. \cite{frohlich}\cite{ctz5}
according to the physical request of locality of the electrons with respect
to the edge excitations. Indeed our projection, when applied to a local
field (namely the electron field in the case of filling factor $\nu =1$),
automatically couples the discrete $Z_{m}$ charge of $U(1)$ with the neutral
sector in order to give rise to a well defined, i. e. single valued,
composite field. Let us also notice that the $m$-electron vertex operator
does not contain any neutral field, so its wave function is realized only by
means of the $c_{X}=1$ charged sector: we deal with a pseudoparticle with
electric charge $m$ and magnetic charge $2pm+1$. The above construction has
been generalized to the torus topology as well \cite{cgm3}, confirming the
picture just outlined for the spectrum of excitations of a QHF at Jain
fillings.

\section{General isomorphism between NCFTs and CFTs: Morita duality at work}

In this Section we exploit the crucial issue of Morita equivalence on
noncommutative tori for rational values of the noncommutativity parameter $%
\theta $. That allows us to build up a general isomorphism between NCFTs and
CFTs on the ordinary space. We will make an explicit reference to the $m$%
-reduced theory describing a QHF at Jain fillings, recalled in
Section 2. We obtain two main results: i) from a theoretical
perspective, a new characterization of the $m$-reduction procedure
is derived, as the image in the ordinary space of Morita duality;
ii) from a more applicative perspective, a new relationship
emerges between noncommutativity and QHF physics.

The Morita equivalence \cite{morita}\cite{morita1} is an isomorphism between
noncommutative algebras that conserves all the modules and their associated
structures. Let us consider an $U(N)$ NCFT defined on the noncommutative
torus ${\rm T}_{\theta }^{2}$ and, for simplicity, of radii $R$. The
coordinates satisfy the commutation rule $[x_{1},x_{2}]=i\theta $ \cite
{ncft1}. In such a simple case the Morita duality is represented by the
following $SL(2,Z)$ action on the parameters:
\begin{equation}
\theta ^{^{\prime }}=\frac{a\theta +b}{c\theta +d};\text{ \ \ \ \ \ \ \ \ }%
R^{^{\prime }}=\left| c\theta +d\right| R,  \label{morita1}
\end{equation}
where $a,b,c,d$ are integers and $ad-bc=1$.

For rational values of the non commutativity parameter, $\theta =-\frac{b}{a}
$, so that $c\theta +d=\frac{1}{a}$, the Morita transformation (\ref{morita1}%
) sends the NCFT to an ordinary one with $\theta ^{^{\prime }}=0$ and
different radius $R^{^{\prime }}=\frac{R}{a}$, involving in particular a
rescaling of the rank of the gauge group \cite{moreno1}\cite{moreno2}\cite
{schiappa}\cite{alvarez1}. Indeed the dual theory is a twisted $%
U(N^{^{\prime }})$ theory with $N^{^{\prime }}=aN$. The classes of
$\theta ^{^{\prime }}=0$ theories are parametrized by an integer
$m$, so that for any $m$ there is a finite number of abelian
theories which are related by a subset of the transformations
given in Eq. (\ref{morita1}). Such observation is a crucial one
and allows us to make the following fundamental statement.

The $m$-reduction technique applied to the QHF at Jain fillings
($\nu =\frac{m}{2pm+1}$) can be viewed as the image of the Morita
map (characterized by $a=2p(m-1)+1$, $b=2p$, $c=m-1$, $d=1$)
between the two NCFTs with $\theta =1$ and $\theta
=2p+\frac{1}{m}$ respectively and corresponds to the Morita map in
the ordinary space. The $\theta =1$ theory is an $U\left( 1\right)
_{\theta =1}$ NCFT while the mother CFT is an ordinary $U\left(
1\right) $ theory; furthermore, when the $U\left( 1\right)
_{\theta =2p+\frac{1}{m}}$ NCFT is considered, its Morita dual CFT has $%
U\left( m\right) $ symmetry. As a consequence, the following correspondence
Table between the NCFTs and the ordinary CFTs is established:
\begin{equation}
\begin{array}{ccc}
& \text{Morita} &  \\
U\left( 1\right) _{\theta =1} & \rightarrow  & U\left( 1\right) _{\theta =0}
\\
& \left( a=1,b=-1,c=0,d=1\right)  &  \\
\text{Morita}\downarrow \left( a,b,c,d\right)  &  & m-\text{reduction}%
\downarrow  \\
& \text{Morita} &  \\
U\left( 1\right) _{\theta =2p+\frac{1}{m}} & \rightarrow  & U\left( m\right)
_{\theta =0} \\
& \left( a=m,b=-2pm-1,c=1-m,d=2p\left( m-1\right) +1\right)  &
\end{array}
\label{morita2}
\end{equation}
This is the main result of the paper.

For more general commutativity parameters $\theta =\frac{q}{m}$ such a
correspondence can be easily extended. Indeed the action of the $m$%
-reduction procedure on the number $q$ doesn't change the central charge of
the CFT under study but modifies the compactification radius of the charged
sector \cite{cgm1}\cite{cgm3}. Nevertheless in this paper we are interested
to the action of the Morita map on the denominator of the parameter $\theta $
which has interesting consequences on noncommutativity, so in the following
we will concentrate on such an issue. The generalization of the Morita map
to different rational noncommutativity parameters, and at the same time to
the physics of QHF at different filling factors, will be the subject of a
future publication \cite{nos1}.

In order to show that the $m$-reduction technique applied to the
QHF at Jain fillings is the image of the Morita map between the
two NCFTs with $\theta =1$ and $\theta =2p+\frac{1}{m}$
respectively and corresponds to the Morita map in the ordinary
space it is enough to show how the twisted boundary conditions on
the neutral fields of the $m$-reduced theory (see Section 2) arise
as a consequence of the noncommutative nature of the $U\left(
1\right) _{\theta =2p+\frac{1}{m}}$ NCFT.

In order to carry out this program let us recall that an associative algebra
of smooth functions over the noncommutative two-torus ${\rm T}_{\theta }^{2}$
can be realized through the Moyal product ($\left[ x_{1},x_{2}\right]
=i\theta $):
\begin{equation}
f\left( x\right) \ast g\left( x\right) =\left. \exp \left( \frac{i\theta }{2}%
\left( \partial _{x_{1}}\partial _{y_{2}}-\partial
_{x_{2}}\partial _{y_{1}}\right) \right) f\left( x\right)  g\left(
y\right) \right| _{y=x}.  \label{moyal1}
\end{equation}
It is convenient to decompose the elements of the algebra, i. e. the fields,
in their Fourier components. However a general field operator $\Phi $
defined on a torus can have different boundary conditions associated to any
of the compact directions. For the torus we have four different
possibilities:
\begin{equation}
\begin{array}{cc}
\Phi \left( x_{1}+R,x_{2}\right) =e^{2\pi i\alpha _{1}}\Phi \left(
x_{1},x_{2}\right) , & \Phi \left( x_{1},x_{2}+R\right) =e^{2\pi i\alpha
_{2}}\Phi \left( x_{1},x_{2}\right) ,
\end{array}
\label{bc1}
\end{equation}
where $\alpha _{1}$ and $\alpha _{2}$ are the boundary parameters. The
Fourier expansion of the general field operator $\Phi _{\overrightarrow{%
\alpha }}$ with boundary conditions $\overrightarrow{\alpha }=\left( \alpha
_{1},\alpha _{2}\right) $ takes the form:
\begin{equation}
\Phi _{\overrightarrow{\alpha }}=\sum_{\overrightarrow{n}}\Phi ^{%
\overrightarrow{n}}U_{\overrightarrow{n}+\overrightarrow{\alpha }}
\label{fexp1}
\end{equation}
where we define the generators as
\begin{equation}
U_{\overrightarrow{n}}\equiv \exp \left( 2\pi i\frac{\overrightarrow{n}\cdot
\overrightarrow{x}}{R}\right) .  \label{fexp2}
\end{equation}
They give rise to the following Moyal commutator:
\begin{equation}
\left[ U_{\overrightarrow{n}+\overrightarrow{\alpha }},U_{\overrightarrow{%
n^{\prime }}+\overrightarrow{\alpha ^{\prime }}}\right] =-2i\sin \left(
\frac{2\pi ^{2}\theta }{R^{2}}\left( \overrightarrow{n}+\overrightarrow{%
\alpha }\right) \wedge \left( \overrightarrow{n^{\prime }}+\overrightarrow{%
\alpha ^{\prime }}\right) \right) U_{\overrightarrow{n}+\overrightarrow{%
n^{\prime }}+\overrightarrow{\alpha }+\overrightarrow{\alpha ^{\prime }}},
\label{fexp3}
\end{equation}
where $\overrightarrow{p}\wedge \overrightarrow{q}=\varepsilon
_{ij}p_{i}q_{j}$.

When the noncommutativity parameter $\theta $ takes the rational value $%
\theta =\frac{2q}{m}\frac{R^{2}}{2\pi }$, being $q$ and $m$ relatively prime
integers, the infinite-dimensional algebra generated by the $U_{%
\overrightarrow{n}+\overrightarrow{\alpha }}$ breaks up into equivalence
classes of finite dimensional subspaces. Indeed the elements $U_{m%
\overrightarrow{n}}$ generate the center of the algebra and that makes
possible for the momenta the following decomposition:
\begin{equation}
\overrightarrow{n^{\prime }}+\overrightarrow{\alpha }=m\overrightarrow{n}+%
\overrightarrow{n},\text{ \ \ \ \ }0\leq n_{1},n_{2}\leq m-1.  \label{fexp4}
\end{equation}
The whole algebra splits into equivalence classes classified by all the
possible values of $m\overrightarrow{n}$, each class being a subalgebra
generated by the $m^{2}$ functions $U_{\overrightarrow{n}+\overrightarrow{%
\alpha }}$ which satisfy the relations
\begin{equation}
\left[ U_{\overrightarrow{n}+\overrightarrow{\alpha }},U_{\overrightarrow{%
n^{\prime }}+\overrightarrow{\alpha ^{\prime }}}\right] =-2i\sin
\left( \frac{\pi q}{m}\left(
\overrightarrow{n}+\overrightarrow{\alpha }\right)
\wedge \left( \overrightarrow{n^{\prime }}+\overrightarrow{\alpha ^{\prime }}%
\right) \right) U_{\overrightarrow{n}+\overrightarrow{n^{\prime }}+%
\overrightarrow{\alpha }+\overrightarrow{\alpha ^{\prime }}}.  \label{fexp5}
\end{equation}
The algebra (\ref{fexp5}) is isomorphic to the (complexification of the) $%
U\left( m\right) $ algebra, whose general $m$-dimensional representation can
be constructed by means of the following ''shift'' and ''clock'' matrices
\cite{matrix1}\cite{matrix2}\cite{matrix3}:
\begin{equation}
Q=\left(
\begin{array}{cccc}
1 &  &  &  \\
& \varepsilon  &  &  \\
&  & \ddots  &  \\
&  &  & \varepsilon ^{m-1}
\end{array}
\right) ,\text{ \ \ \ \ \ \ }P=\left(
\begin{array}{cccc}
0 & 1 &  & 0 \\
& \cdots  &  &  \\
&  & \vdots  & 1 \\
1 &  &  & 0
\end{array}
\right) ,  \label{fexp6}
\end{equation}
being $\varepsilon =\exp (\frac{2\pi iq}{m})$. So the matrices $J_{%
\overrightarrow{n}}=\varepsilon ^{n_{1}n_{2}}Q^{n_{1}}P^{n_{2}}$, $%
n_{1},n_{2}=0,...,m-1$, generate an algebra isomorphic to (\ref{fexp5}):
\begin{equation}
\left[ J_{\overrightarrow{n}},J_{\overrightarrow{n^{\prime }}}\right]
=-2i\sin \left( \pi \frac{q}{m}\overrightarrow{n}\wedge \overrightarrow{%
n^{\prime }}\right) J_{\overrightarrow{n}+\overrightarrow{n^{\prime }}}.
\label{fexp7}
\end{equation}
Thus the following Morita mapping has been realized between the Fourier
modes defined on a noncommutative torus and functions taking values on $%
U\left( m\right) $ but defined on a commutative space:
\begin{equation}
\exp \left( 2\pi i\frac{\left( \overrightarrow{n}+\overrightarrow{\alpha }%
\right) \cdot \widehat{\overrightarrow{x}}}{R}\right) \longleftrightarrow
\exp \left( 2\pi i\frac{\left( \overrightarrow{n}+\overrightarrow{\alpha }%
\right) \cdot \overrightarrow{x}}{R}\right) J_{\overrightarrow{n}+%
\overrightarrow{\alpha }}.  \label{fexp8}
\end{equation}
As a consequence a mapping between the fields $\Phi _{\overrightarrow{\alpha
}}$ is generated as follows. Let us focus, for simplicity, on the case $q=1$
which leads for the momenta to the decomposition $\overrightarrow{n}=m%
\overrightarrow{n}+\overrightarrow{j}$, with \ \ \ \ $0\leq j_{1},j_{2}\leq m
$. The general field operator $\Phi _{\overrightarrow{\alpha }}$ on the
noncommutative torus ${\rm T}_{\theta }^{2}$ with boundary conditions $%
\overrightarrow{\alpha }$ can be written in the form:
\begin{equation}
\Phi _{\overrightarrow{\alpha }}=\sum_{\overrightarrow{n}}\exp \left( 2\pi im%
\frac{\overrightarrow{n}\cdot \overrightarrow{x}}{R}\right) \sum_{%
\overrightarrow{j}=0}^{m^{\prime }-1}\Phi ^{\overrightarrow{n},%
\overrightarrow{j}}U_{\overrightarrow{j}+\overrightarrow{\alpha }}.
\label{fexp9}
\end{equation}
By using Eq. (\ref{fexp8}) we obtain the Morita correspondence between
fields as:
\begin{equation}
\Phi _{\overrightarrow{\alpha }}\longleftrightarrow \Phi =\sum_{%
\overrightarrow{j}=0}^{m^{\prime }-1}\chi ^{\left( \overrightarrow{j}\right)
}J_{\overrightarrow{j}+\overrightarrow{\alpha }},  \label{fexp10}
\end{equation}
where we have defined:
\begin{equation}
\chi ^{\left( \overrightarrow{j}\right) }=\exp \left( 2\pi i\frac{\left(
\overrightarrow{j}+\overrightarrow{\alpha }\right) \cdot \overrightarrow{x}}{%
R}\right) \sum_{\overrightarrow{n}}\Phi ^{\overrightarrow{n},\overrightarrow{%
j}}\exp \left( 2\pi im\frac{\overrightarrow{n}\cdot \overrightarrow{x}}{R}%
\right) .  \label{fexp11}
\end{equation}
The field $\Phi $ is defined on the dual torus with radius $R^{\prime }=%
\frac{R}{m^{\prime }}$ and satisfies the boundary conditions:
\begin{equation}
\begin{array}{cc}
\Phi \left( \theta +R^{\prime },x_{2}\right) =\Omega _{1}^{+}\cdot \Phi
\left( \theta ,x_{2}\right) \cdot \Omega _{1}, & \Phi \left( \theta
,x_{2}+R^{\prime }\right) =\Omega _{2}^{+}\cdot \Phi \left( \theta
,x_{2}\right) \cdot \Omega _{2},
\end{array}
\label{fexp12}
\end{equation}
with
\begin{equation}
\Omega _{1}=P^{b},\text{ \ \ \ \ \ \ }\Omega _{2}=Q^{1/q},  \label{fexp13}
\end{equation}
where $b$ is an integer satisfying $am-bq=1$. While the field components $%
\chi ^{\left( \overrightarrow{j}\right) }$ satisfy the following twisted
boundary conditions:
\begin{equation}
\begin{array}{c}
\chi ^{\left( \overrightarrow{j}\right) }\left( \theta +R^{\prime
},x_{2}\right) =e^{2\pi i\left( j_{1}+\alpha _{1}\right) /m}\chi ^{\left(
\overrightarrow{j}\right) }\left( \theta ,x_{2}\right)  \\
\chi ^{\left( \overrightarrow{j}\right) }\left( \theta ,x_{2}+R^{\prime
}\right) =e^{2\pi i\left( j_{2}+\alpha _{2}\right) /m}\chi ^{\left(
\overrightarrow{j}\right) }\left( \theta ,x_{2}\right)
\end{array}
,  \label{fexp14}
\end{equation}
that is
\begin{equation}
\left( \frac{j_{1}+\alpha _{1}}{m},\frac{j_{2}+\alpha _{2}}{m}\right) ,\text{
\ \ \ \ }j_{1}=0,...,m-1,\text{ \ \ \ \ }j_{2}=0,...,m-1.  \label{fexp15}
\end{equation}
Let us observe that $\overrightarrow{j}=\left( 0,0\right) $ is the
trace degree of freedom which can be identified with the $U(1)$
component of the matrix valued field or the charged component
within the $m$-reduced theory of the QHF at Jain fillings
introduced in Section 2. We infer that only the integer part of
$\frac{n_{i}}{m}$ should really be thought of as the momentum. The
commutative torus is smaller by a factor $m\times m$ than the
noncommutative one; in fact upon this rescaling also the ''density
of degrees of freedom'' is kept constant as now we are dealing
with $m\times m$ matrices instead of scalars.

Let us now summarize what we have learnt. When the parameter $\theta $ is
rational we recover the whole structure of the noncommutative torus and
recognize the twisted boundary conditions which characterize the neutral
fields (\ref{phi}) of the $m$-reduced theory as the consequence of the
Morita mapping of the starting NCFT ($U\left( 1\right) _{\theta =2p+\frac{1}{%
m}}$ in our case) on the ordinary commutative space. Indeed $\chi ^{\left(
0,0\right) }$ corresponds to the charged $X$ field while the twisted fields $%
\chi ^{\left( \overrightarrow{j}\right) }$ with $\overrightarrow{j}\neq
\left( 0,0\right) $ should be identified with the neutral ones (\ref{phi}).
Therefore the $m$-reduction technique can be viewed as a realization of the
Morita mapping between NCFTs and CFTs on the ordinary space, as sketched in
the Table (\ref{morita2}). In the following Sections we further clarify such
a correspondence by making an explicit reference to the QHF physics at Jain
fillings. In particular we will recognize the generalized magnetic
translations as a realization of the Moyal algebra defined in Eq. (\ref
{fexp5}).

\section{Noncommutative geometry from $m$-reduction}

In this Section we study in detail how the $m$-reduction technique is the
image on the ordinary space of the Morita duality. In order to shed further
light on such a correspondence let us follow the main steps of the procedure
and extract the properties which are relevant for noncommutativity.

As shown in Section 2, the modes of the $m$-reduced theory on the plane
naturally split into two subsets: the first, built up with indices which
divide by the integer $m$, is the new affinization of the operator algebra
while the second, built up with the modes indexed by $j=1,...,m-1$, is
defined on a discrete version of the circle and gives rise to the new zero
modes. That corresponds to the following decomposition of the $\left|
z\right| =1$ circle (parametrized by the phase variable $\sigma $):
\begin{equation}
S_{m}^{1}\times S^{1},  \label{circle1}
\end{equation}
where $S_{m}^{1}=\frac{2\pi i}{m}Z_{m}$ is a discrete version of the circle.
Within a first quantization scheme we may introduce the conjugate variable $%
\widetilde{\sigma }=i\partial _{\sigma }$ which can be viewed as a momentum
and takes values in the dual lattice $Z_{m}\times Z$. The following usual
commutation relation between conjugate variables holds:
\begin{equation}
\left[ \widetilde{\sigma },\sigma \right] =i,  \label{circle2}
\end{equation}
which makes transparent the identification with the physics of a
two-dimensional QHF. Indeed the parameter space is the discrete two-torus $%
{\rm T}^{2}=S_{m}^{1}\times Z_{m}$, representing positions and momenta of a
particle on a circle made of $m$ points given by the angles $0,\frac{2\pi }{m%
},...,\frac{2\pi \left( m-1\right) }{m}$. A realization of the generators of
the $su(m)$ algebra on such a space (${\rm T}^{2}$) can be easily obtained
in terms of the two following fundamental translations:
\begin{equation}
\begin{array}{cc}
J_{j_{1},0}=z^{j_{1}}, & J_{0,j_{2}}=\varepsilon ^{j_{2}\widetilde{\sigma }}
\end{array}
,  \label{circles3}
\end{equation}
where $\varepsilon =e^{i\frac{2\pi }{m}}$. They correspond to the
position and the momentum respectively and define a projective
representation of the abelian group of translations on the
discrete two-torus ${\rm T}^{2}$. Indeed the $m^{2}-1$ general
operators:
\begin{equation}
\begin{array}{cc}
J_{j_{1},j_{2}}=\varepsilon ^{\frac{j_{1}j_{2}}{2}}z^{j_{1}}\varepsilon
^{j_{2}\widetilde{\sigma }}, &
\begin{array}{c}
j_{1},j_{2}=0,...,m-1 \\
\left( j_{1},j_{2}\right) \neq \left( 0,0\right)
\end{array}
\end{array}
\label{circles4}
\end{equation}
satisfy the commutation relations of the $su(m)$ algebra.

Such a realization is relevant for the study of the properties of
the QHF wave function and can be obtained by means of the
restriction to the discrete space of the symmetry algebra
$W_{1+\infty }$ of area preserving diffeomorphisms of the
two-torus in the continuum. So, let us summarize its basic
properties which will be useful in the following. Let us start by
recalling that the elements of the modular group $SL(2,Z)$ can be
defined on the discrete two-torus ${\rm T}^{2}$ with rational
coordinates sharing the
same denominator $m$: $\left( n,l\right) \equiv \left( \frac{j_{1}}{m},\frac{%
j_{2}}{m}\right) \in {\rm T}^{2}$, where $j_{1},j_{2},m\in Z$; in
particular the $mod 1$ action becomes a $mod m$ action on the
equivalent two-torus $\left( n,l\right) \in m{\rm T}^{2}$. That
allows us to define the relevant discrete subgroup of the group of
area preserving diffeomorphisms of the continuous two-torus as the
coset $SL(2,Z_{m})=\frac{SL(2,Z)}{\Gamma
\left( m\right) }$, where $\Gamma \left( m\right) $ is the set of matrices $%
g\in SL(2,Z)$ such that $g=\pm I$ $mod m$. The main property is
the following.

For $m$ prime and integer, the generators (\ref{circles4}) of $su(m)$ are a
single irreducible representation under $SL(2,Z_{m})$. Indeed the action of $%
SL(2,Z_{m})$ on such generators, for instance $J_{1,0}$, is defined as:
\begin{equation}
J_{j_{1},j_{2}}=V\left( \overline{g}\right) J_{1,0}V^{-1}\left( \overline{g}%
\right) ,  \label{circles5}
\end{equation}
where the matrix $\overline{g}\in SL(2,Z_{m})$ ($\det
\overline{g}=1 mod m$) acts upon the vector $\left( 1,0\right) $
as:
\begin{equation}
\left(
\begin{array}{c}
j_{1} \\
j_{2}
\end{array}
\right) =\left(
\begin{array}{cc}
a & b \\
c & d
\end{array}
\right) \left(
\begin{array}{c}
1 \\
0
\end{array}
\right) .  \label{circles6}
\end{equation}
Thus, the subgroup $S_{m}$, built of matrices of the kind
\begin{equation}
\left(
\begin{array}{cc}
1 & b \\
0 & 1
\end{array}
\right) ,  \label{circles7}
\end{equation}
is the stability group for the vector $\left( 1,0\right) $, so that the
number of $\frac{SL(2,Z_{m})}{S_{m}}$ equivalence classes coincides with $%
m^{2}-1$, i. e. the number of generators of the $su(m)$ algebra.

For $m$ arbitrary, such that $m=\prod_{i=1}^{s}m_{i}^{n_{i}}$ and $m_{i}$
prime, we get $SL(2,Z_{m})=\otimes _{i=1}^{s}SL(2,Z_{m_{i}^{n_{i}}})$ and $%
\sum_{i=1}^{s}n_{i}$ independent orbits are obtained.

The main result is that all the possible states of the QHF on the discrete
two-torus can be classified in terms of the coset $SL(2,Z_{m})=\frac{SL(2,Z)%
}{\Gamma \left( m\right) }$. As a consequence, the ordinary space of CFTs
obtained through the Morita map shown in Table (\ref{morita2}) acquires such
a structure and the conformal fields (\ref{fexp11}) (i. e. (\ref{phi}))
defined on it gain twisted boundary conditions (see Section 3).

\section{Generalized magnetic translations as a realization of Moyal algebra}

In this Section we will make explicit reference to the $m$-reduced theory
for a QHF at Jain fillings and to the issue of GMT on a torus in order to
identify in such a context a realization of the Moyal algebra.

Let us recall that all the relevant topological effects in the QHF
physics, such as the degeneracy of the ground state wave function
on a manifold with non trivial topology, the derivation of the
Hall conductance $\sigma _{H}$ as a topological invariant and the
relation between fractional charge and statistics of anyon
excitations \cite{top1}\cite{top2}, can be made very transparent
by using the invariance properties of the wave functions under a
finite subgroup of the magnetic translation group for a $N_{e}$
electrons system. Indeed their explicit expressions as the
Verlinde operators \cite {top3}, which generate the modular
transformations in the $c=1$ CFT, are taken as a realization of
topological order of the system under study \cite{wen}. In
particular the magnetic translations built so far
\cite{cristofano1} act on the characters associated to the highest
weight states which represent the charged statistical particles,
the anyons or the electrons.

In our CFT representation of the QHF at Jain fillings \cite{cgm1}\cite{cgm3}
we shall see that the Moyal algebra defined in Eq. (\ref{fexp5}) has a
natural and beautiful realization in terms of GMT. We refer to them as
generalized ones because the usual magnetic translations act on the charged
content of the one point functions \cite{cristofano1}. Instead, in our TM
model for the QHF (see Section 2 and Refs. \cite{cgm3}, \cite{cgm4}) the
primary fields (and then the corresponding characters within the torus
topology) appear as composite field operators which factorize in a charged
as well as a neutral part. Further they are also coupled by the discrete
symmetry group $Z_{m}$. Then, in order to show that the characters of the
theory are closed under magnetic translations, we need to generalize them in
such a way that they will appear as operators with two factors, acting on
the charged and on the neutral sector respectively.

Let us also recall that the incompressibility of the quantum Hall fluid
naturally leads to a $W_{1+\infty }$ dynamical symmetry \cite{BS, ctz}.
Indeed, if one considers a droplet of a quantum Hall fluid, it is evident
that the only possible area preserving deformations of this droplet are the
waves at the boundary of the droplet, which describe the deformations of its
shape, the so called edge excitations. These can be well described by the
infinite generators $W_{m}^{n+1}$ of $W_{1+\infty }$ of conformal spin $%
(n+1) $, which are characterized by a mode index $m\in Z$ and satisfy the
algebra:
\begin{equation}
\left[ W_{m}^{n+1},W_{m^{\prime }}^{n^{\prime }+1}\right]
=(n^{\prime }m-nm^{\prime })W_{m+m^{\prime }}^{n+n^{\prime
}}+q(n,n^{\prime },m,m^{\prime })W_{m+m^{\prime }}^{n+n^{\prime
}-2}+...+d(n,m)c\,\delta ^{n,n^{\prime }}\delta _{m+m^{\prime
}=0},  \label{walgebras}
\end{equation}
where the structure constants $q$ and $d$ are polynomials of their
arguments, $c$ is the central charge, and dots denote a finite number of
similar terms involving the operators $W_{m+m^{\prime }}^{n+n^{\prime }-2l}$
\cite{BS, ctz}. Such an algebra contains an Abelian $\widehat{U}(1)$ current
for $n=0$ and a Virasoro algebra for $n=1$ with central charge $c$. It
encodes the local properties which are imposed by the incompressibility
constraint and realizes the allowed edge excitations \cite{Cappelli}.
Nevertheless algebraic properties do not include topological properties
which are also a consequence of incompressibility. In order to take into
account the topological properties we have to resort to finite magnetic
translations which encode the large scale behaviour of the QHF.

Let us consider a magnetic translation of step $(n=n_{1}+in_{2},\overline{n}%
=n_{1}-in_{2})$ on a sample with coordinates $x_{1},x_{2}$ and define the
corresponding generators $T^{n,\overline{n}}$ as:
\begin{equation}
T^{n,\overline{n}}=e^{-\frac{B}{4}n\overline{n}}e^{\frac{1}{2}nb^{+}}e^{%
\frac{1}{2}\overline{n}b},  \label{tr1}
\end{equation}
where $b^{+}=i\partial _{\overline{\omega }}-i\frac{B}{2}\omega $, $%
b=i\partial _{\omega }+i\frac{B}{2}\overline{\omega }$, $\omega $ is a
complex coordinate ($\overline{\omega }$ being its conjugate) and $B$ is the
transverse magnetic field. They satisfy the relevant property:
\begin{equation}
T^{n,\overline{n}}T^{m,\overline{m}}=q^{-\frac{n\times m}{4}}T^{n+m,}{}^{%
\overline{n}+\overline{m}},  \label{tr2}
\end{equation}
where $q$ is a root of unity.

Furthermore, it can be easily shown that they admit the following expansion
in terms of the generators $W_{k-1}^{l-1}$ of the $W_{1+\infty }$ algebra:
\begin{equation}
T^{n,\overline{n}}=e^{-\frac{B}{4}n\overline{n}}\sum_{k,l=0}^{\infty }\left(
-\right) ^{l}\frac{n^{k}}{2^{k}n}\frac{\overline{n}^{l}}{2^{l}\overline{n}}%
W_{k-1}^{l-1},  \label{tr3}
\end{equation}
where now the local $W_{1+\infty }$ symmetry and the global topological
properties are much more evident because the coefficients in the above
series depend on the topology of the sample.

Within our $m$-reduced theory for a QHF at Jain fillings \cite{cgm1}\cite
{cgm3}, introduced in Section 2, it can be shown that also magnetic
translations of step $(n,\overline{n})$ decompose into equivalence classes
and can be factorized into a group, with generators $T_{C}^{n,\overline{n}}$%
, which acts only on the charged sector as well as a group, with generators $%
T_{S}^{j_{i},\overline{j}}$, acting only on the neutral sector \cite{nos2}.
The presence of the transverse magnetic field $B$ reduces the torus to a
noncommutative one and the flux quantization induces rational values of the
noncommutativity parameter $\theta $. As a consequence the neutral magnetic
translations realize a projective representation of the $su\left( m\right) $
algebra generated by the elementary translations:
\begin{equation}
J_{a,b}=e^{-2\pi i\frac{ab}{m}}T_{S}^{a,0}T_{S}^{0,b};\text{ \ \ \ \ \ \ }%
a,b=1,...,m,  \label{tr4}
\end{equation}
which satisfy the commutation relations:
\begin{equation}
\left[ J_{a,b},J_{\alpha ,\beta }\right] =-2i\sin \left( \frac{2\pi }{m}%
\left( a\beta -b\alpha \right) \right) J_{a+\alpha ,b+\beta }.  \label{tr5}
\end{equation}
Let us notice that the $su\left( m\right) $ generators introduced in Eq. (%
\ref{circles4}) coincide with the GMT ones (\ref{tr4}); in this context Eq. (%
\ref{circles5}) gives the action of the modular group on the GMT generators
and clarifies the relevance of the hidden $su\left( m\right) $ algebra in
classifying the topological excitations of the QHF.

The GMT operators above defined (see Eqs. (\ref{tr1}) and (\ref{tr4})) are a
realization of the Moyal operators introduced in Eq. (\ref{fexp2}) and the
algebra defined by Eq. (\ref{tr5}) is isomorphic to the Moyal algebra given
in Eqs. (\ref{fexp5}) and (\ref{fexp7}). Such operators generate the
residual symmetry of the $m$-reduced CFT which is Morita equivalent to the
NCFT with rational non commutativity parameter $\theta =2p+\frac{1}{m}$.
Furthermore they can be identified with the $su\left( m\right) $ {\it twist
eaters} matrices $\Gamma _{a}$ which generate the Weyl-t'Hooft algebra \cite
{matrix3}:
\begin{equation}
\Gamma _{a}\Gamma _{b}=e^{2\pi iQ_{ab}/m}\Gamma _{b}\Gamma _{a},
\label{weyl1}
\end{equation}
where $Q$ is the matrix of nonabelian $su(m)$ t'Hooft fluxes through the non
trivial cycles of the torus. In this way the structure of GMTs within the $m$%
-reduced CFT describing the physics of a QHF at Jain fillings
coincides with that of the $\frac{SU\left( m\right) }{Z_{m}}$
sector of a general $U\left( m\right) $ gauge theory. Indeed
noncommutativity makes the $U\left( 1\right) $ and $SU\left(
m\right) $ sectors of the decomposition:
\begin{equation}
U\left( m\right) =U\left( 1\right) \times \frac{SU\left( m\right) }{Z_{m}}
\label{weyl2}
\end{equation}
not decoupled because the $U\left( 1\right) $ {\it photon} interacts with
the $SU\left( m\right) $ {\it gluons }\cite{gauge1}. Finally, the $U(1)$
sector could be identified also with the center of mass coordinate of a
system of $m$ $D$-branes and it represents the interactions of the short
open string excitations on the $D$-branes with the bulk supergravity fields
\cite{ncft3}\cite{ncft4}. For a vanishing background $B$-field, the closed
and open string dynamics decouple and one is left with an effective $%
SU\left( m\right) $ gauge theory, while this is no longer true for $B\neq 0$.

\section{Conclusions and outlooks}

In conclusion, let us summarize what we have learnt and outline the
perspectives opened by this work.

The Morita equivalence gives rise to an isomorphism between noncommutative
algebras but, in the special case of a rational noncommutativity parameter $%
\theta $, one of the isomorphic theories is a commutative one
\cite{moreno1}\cite{moreno2}\cite{schiappa}. In this paper we have
shown by means of the Morita equivalence that a NCFT with $\theta
=2p+\frac{1}{m}$ is mapped to a CFT on an ordinary space. We
identified such a CFT with the $m$-reduced CFT developed in
\cite{cgm1}\cite{cgm3} for a QHF at Jain fillings, whose neutral
fields satisfy twisted boundary conditions. In this way we gave a
meaning to the concept of ''noncommutative conformal field
theory'', as the Morita equivalent version of a CFT defined on an
ordinary space. The image of Morita duality in the ordinary space
is given by the $m$-reduction technique and the Moyal algebra
which reflects the noncommutative nature is realized by means of
GMT. Furthermore the GMTs structure coincides with that of the
Weyl-t'Hooft algebra \cite{matrix3}. It could be interesting to
further elaborate on the relationship between $U\left( m\right) $
gauge theories and our $m$-reduction procedure, which emerges
through Morita duality in correspondence of rational $\theta $
values \cite{nos3}.

As a remarkable result, a new relation between noncommutative spaces and QHF
physics has been established within a perspective which is very different
from the ones developed in the literature in the last years \cite{qhf1}\cite
{qhf2}\cite{qhf3}\cite{qhf4}. In a more general context a picture emerges on
the interplay between noncommutativity and the physics of strongly
correlated electron systems, which is very different from the ones existing
in the literature \cite{ncmanybody} and deserves further investigations.

The generalization of the Morita map to different rational noncommutativity
parameters $\theta $, and at the same time to the physics of QHF at paired
state filling factors, will be the subject of a future publication \cite
{nos1}. That will help us to shed new light also on the connections between
a system of interacting $D$-branes and the physics of quantum Hall fluids
\cite{branehall1}\cite{branehall2}. Recently the twisted CFT approach
provided by $m$-reduction has been successfully applied to Josephson
junction ladders and arrays of non trivial geometry in order to investigate
the existence of topological order and magnetic flux fractionalization in
view of the implementation of a possible solid state qubit protected from
decoherence \cite{noi3,noi4,noi6} as well as to the study of the phase
diagram of the fully frustrated $XY$ model ($FFXY$) on a square lattice \cite
{noi}. So, it could be interesting to further elucidate the topological
properties of Josephson systems with non trivial geometry and the
consequences imposed by the request of space-time noncommutativity by means
of Morita mapping. Finally, the generality of the $m$-reduction technique
allows us extend many of the present results to $D>2$ theories as string and
$M$-theory.

\end{document}